\documentclass[epj]{webofc}
\usepackage[varg]{txfonts}   
\woctitle{MESON2016 - the 14$^\textrm{th}$ International Workshop on 
Meson Production, Properties and Interaction}
\begin{document}
\selectlanguage{english}
\title{MESON2016 -- Concluding Remarks} 


\author{Avraham Gal\inst{1}\fnsep\thanks{\email{avragal@savion.huji.ac.il}}} 

\institute{Racah Institute of Physics, The Hebrew University, Jerusalem 91904, 
Israel} 

\abstract{Several topics presented and discussed at MESON2016 are highlighted, 
\\ including pentaquarks, dibaryons and meson-nuclear bound states.} 
\maketitle
\section{Introduction}
\label{sec:intro}

The scope of topics presented and discussed in MESON2016 is too broad to be 
covered in one concluding talk. I therefore selected a few central 
topics discussed in MESON2016 where my personal involvement helped making some 
meaningful remarks. These topics are Pentaquarks, Dibaryons, and Meson-Nuclear 
Bound States. I apologize to the many speakers whose presentations were not 
mentioned in this concluding talk.

\section{Exotics: remarks on Pentaquarks} 
\label{sec:penta} 

Regarding {\it pentaquarks}, it is appropriate perhaps to note that the 
${\cal S}=-1$ $\Lambda$(1405) resonance, defying a three-quark classification, 
was predicted in 1959 by Dalitz and Tuan as a $\bar KN$ quasibound state 
\cite{Dalitz59} five years {\it before} the term `quark' was transformed by 
Gell-Mann from Fiction to Physics. A recent LQCD calculation confirms its 
$\bar K N$ hadronic structure \cite{Hall15} as opposed to a tightly bound 
genuine pentaquark. Indeed, the $\Lambda$(1405) emerges naturally below 
the $K^-p\,$ threshold in chiral EFT hadronic approaches \cite{Hyodo13}, 
although as shown in Ciepl\'{y}'s talk \cite{Cieply16} the {\it subthreshold} 
$\bar K N$ scattering amplitudes exhibit appreciable model dependence, with 
consequences for $K^-pp$ quasibound-state searches. 

A ${\cal S}=+1$ $\Theta ^+$(1530) pentaquark was claimed more than 10 years 
ago. Recent dedicated experimental searches have failed to confirm it, 
placing instead upper limits on its coupling to the $KN$ channel \cite{E19}. 
It was argued that the $\Theta ^+$ might be formed copiously in nuclei by 
absorption on {\it two} nucleons, e.g. $K^+d\to\Theta^+p$ \cite{GF05,GF06} 
thereby resolving a long-standing puzzle, discussed in Friedman's 
talk \cite{Friedman16}, regarding the size and $A$ dependence of 
$K^+$ nuclear cross sections at low energies. 

\begin{figure}[htb] 
\centering 
\includegraphics[width=0.7\textwidth,height=5.0cm]{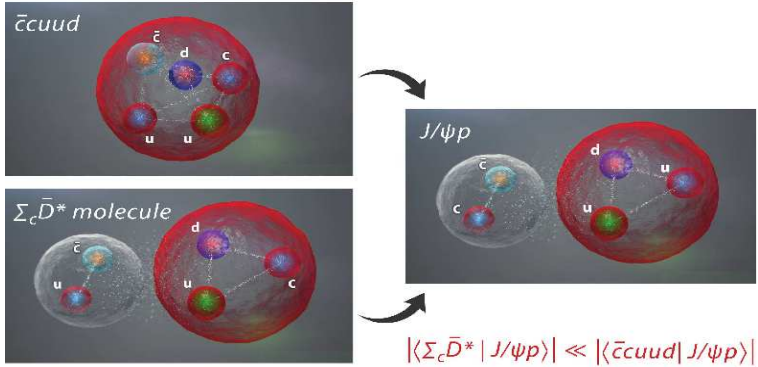} 
\caption{Left: two alternative visual descriptions of the LHCb hidden-charm 
pentaquark $P_c$(4550) as a tightly bound ${\bar c}cuud$ pentaquark or as a 
$\Sigma_c {\bar D}^{\,\ast}$ hadronic molecule. Right: $P_c(4550)\to J/\psi+p$ 
decay channel which for a ${\bar c}cuud$ pentaquark implies a considerably 
larger width than reported. Figure adapted from Karliner's talk.} 
\label{fig:marek} 
\end{figure}

The recent LHCb discovery of hidden-charm structures \cite{LHCb} has led 
to several serious attempts to interpret these in terms of pentaquark(s). 
As argued in Karliner's talk the relatively small width of order 40~MeV 
for $P_c$(4550) supports a $\Sigma_c{\bar D}^{\,\ast}$ hadronic molecule 
structure of two quark clusters, rather than a tightly bound pentaquark; 
see Fig.~\ref{fig:marek} on next page. This $\Sigma_c {\bar D}^{\,\ast}$ 
hadronic molecule is apparently the lightest of several other predicted 
doubly-heavy hadronic molecules \cite{Karliner15}. It was emphasized 
that this calls for a new rich meson-meson, meson-baryon and baryon-baryon 
heavy-flavor QCD spectroscopy. Other speakers too discussed various aspects 
of heavy-flavor Exotics, demonstrating that no clear consensus has yet been 
reached on this topic. 

\section{Exotics: remarks on Dibaryons} 
\label{sec:dibaryons} 

\subsection{Nonstrange dibaryons} 
\label{sec:S=0} 

\begin{figure}[htb] 
\centering 
\includegraphics[width=0.48\textwidth,height=4.5cm]{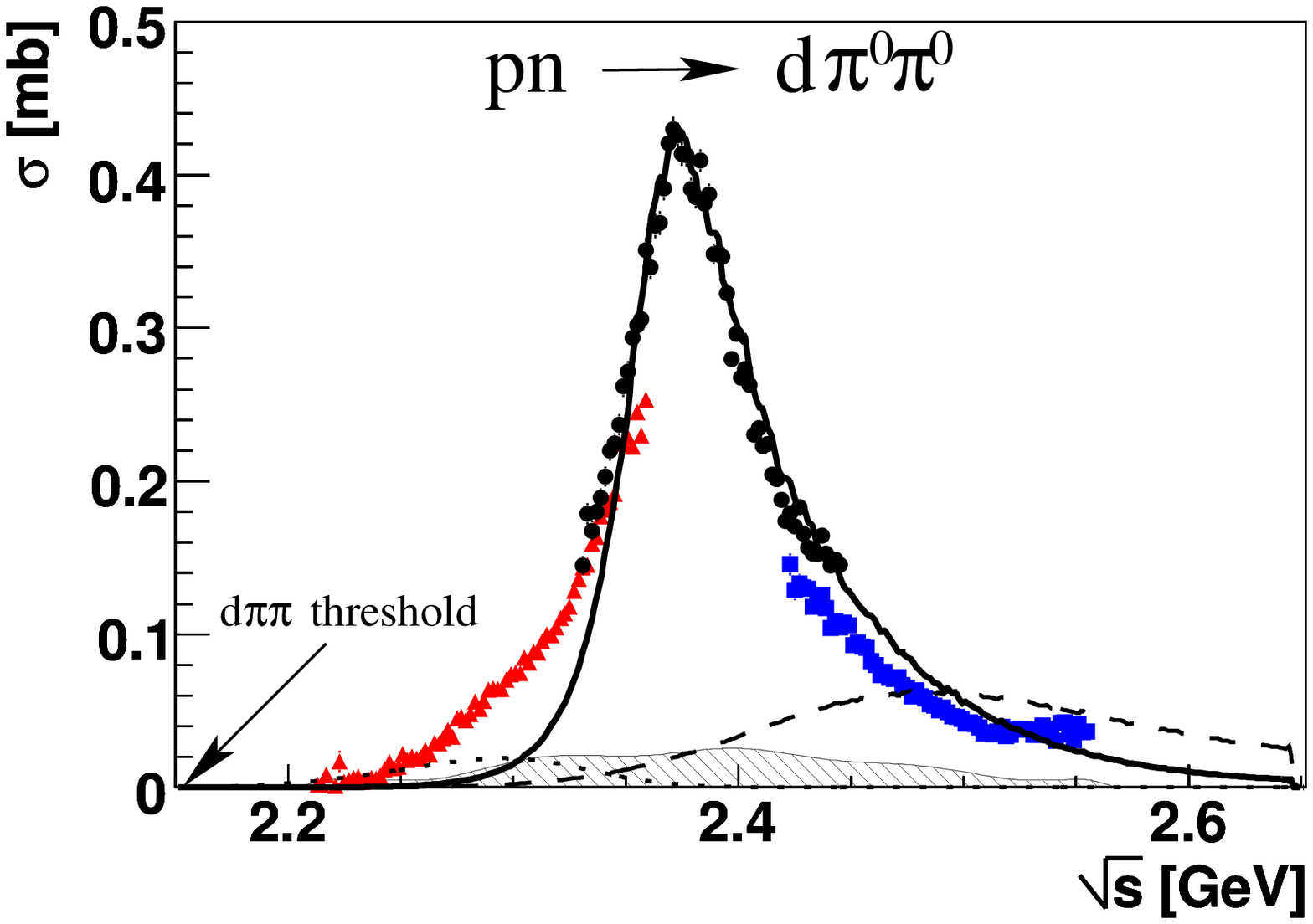} 
\includegraphics[width=0.48\textwidth,height=4.5cm]{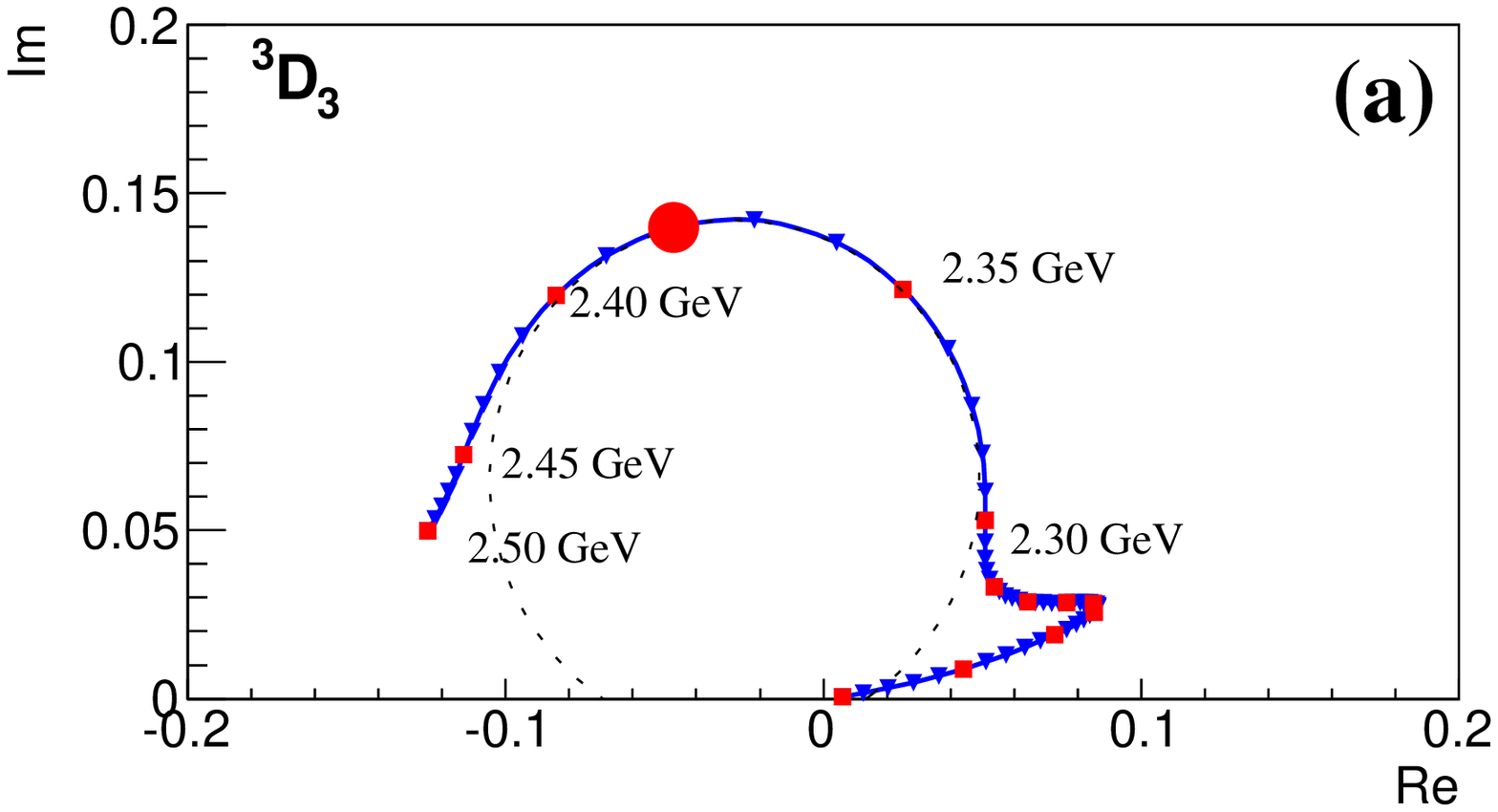} 
\caption{Evidence for the ${\cal D}_{03}(2380)$ dibaryon from WASA-at-COSY. 
Left: from $p+d\to d\pi^0\pi^0+p_{\rm s}$ \cite{COSY11}. Right: from 
the Argand diagram of the ${^3D}_{\,3}$ partial wave in $np$ scattering 
\cite{SAID14}, with full account of the recent measurement of the $np$ 
analyzing power \cite{COSY14}.} 
\label{fig:D03} 
\end{figure}

The only dibaryon for which good experimental evidence exists to date is the 
nonstrange $I=0$ $J^P=3^+$ ${\cal D}_{03}(2380)$, peaking $\approx$85~MeV 
below the $\Delta\Delta$ threshold. The WASA-at-COSY experiments that 
established it, see Fig.~\ref{fig:D03}, were ranked in Wilkin's obituary 
of COSY at MESON2016 a top no. 2 in COSY's impact list. The small width 
of ${\cal D}_{03}(2380)$, $\Gamma\approx 70$~MeV, less than even a single 
$\Delta$ width, was shown (on p.~479 in Ref.~\cite{Gal16}) to follow from 
phase-space and quantum-statistics arguments.  

The large binding energy of ${\cal D}_{03}(2380)$ with respect to $\Delta
\Delta$, exceeding by far the scale of nucleon separation energies in nuclei, 
does not mean it is a deeply bound dibaryon if one recalls the existence 
of a lower two-body channel, $\pi{\cal D}_{12}(2150)$, relative to which 
${\cal D}_{03}(2380)$ resonates. Here, ${\cal D}_{12}(2150)$ stands for 
a near-threshold $N\Delta$ $I=1$ $J^P=2^+$ $\pi NN$ quasibound state. Both 
${\cal D}_{12}$ and ${\cal D}_{03}$, together with their $I\leftrightarrow S$ 
twins ${\cal D}_{21}$ and ${\cal D}_{30}$, were proposed by Dyson and Xuong 
\cite{Dyson64} based on symmetry arguments, and have been considered 
subsequently in numerous quark-based works; and recently also in terms of 
`meson assisted dibaryons' \cite{Gal16} in the hadronic basis \cite{gg13,gg14}. 

\begin{figure}[htb] 
\centering 
\includegraphics[width=0.48\textwidth,height=5.2cm]{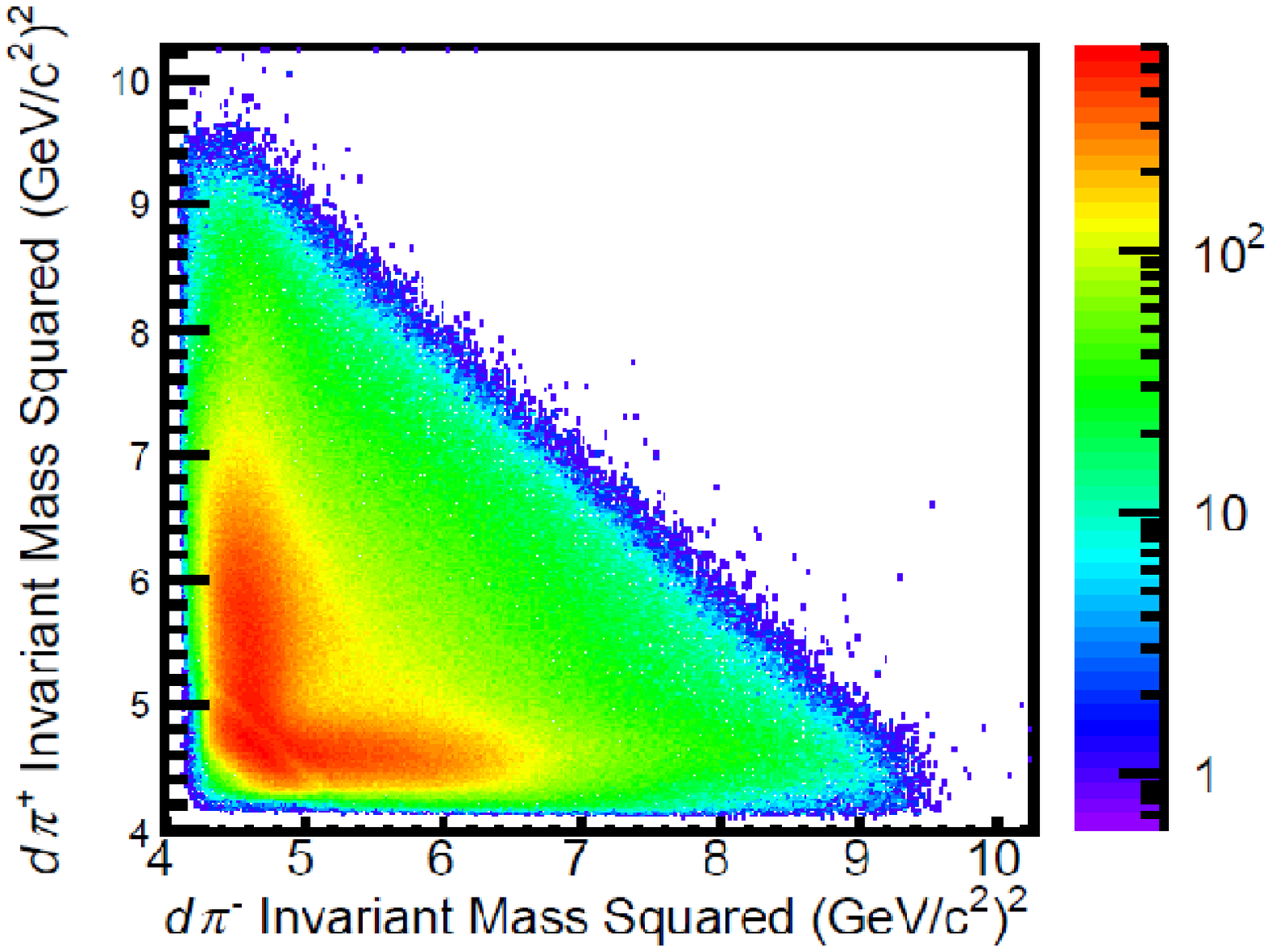} 
\includegraphics[width=0.48\textwidth,height=5cm]{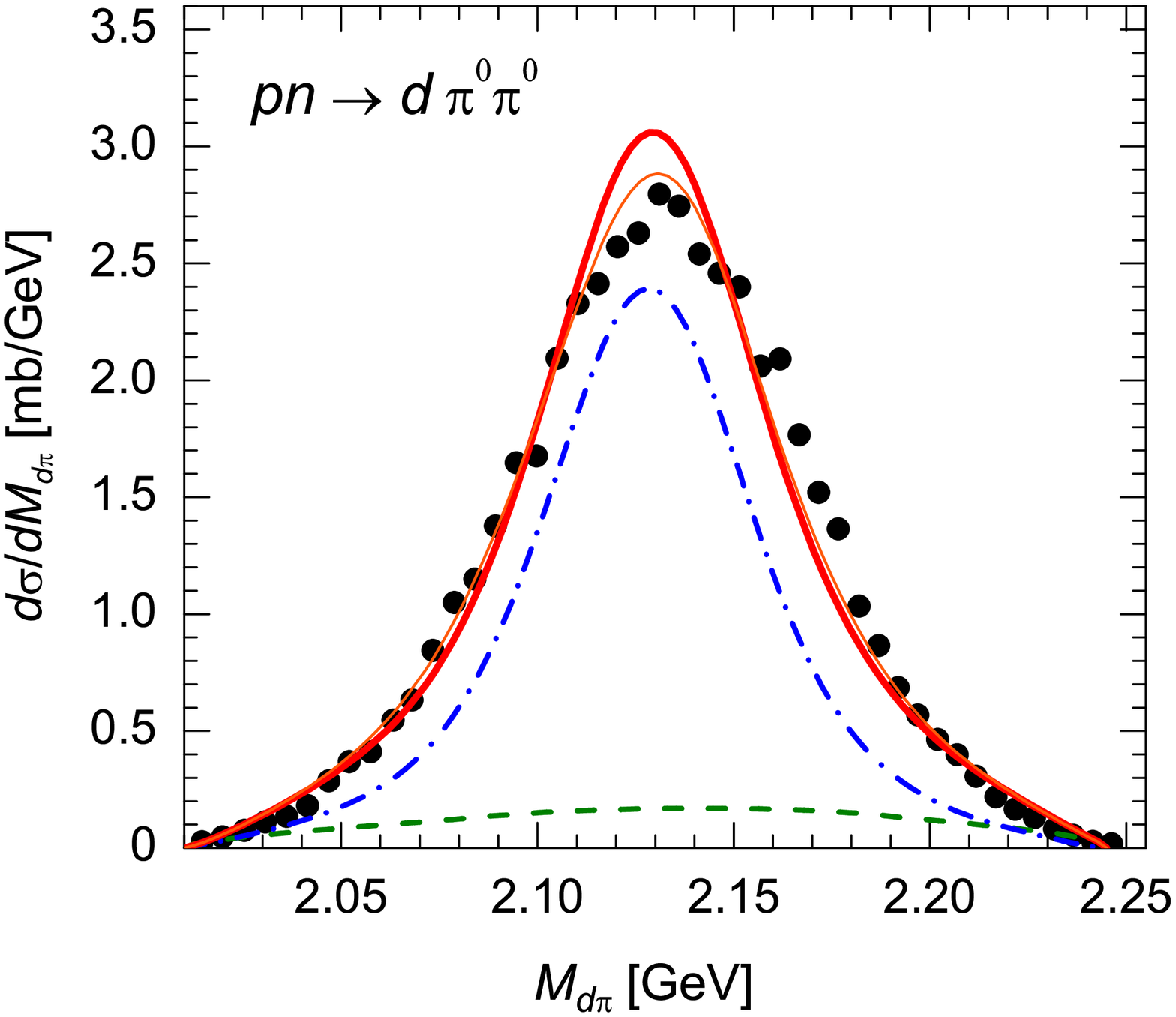}  
\caption{Left: ${\cal D}_{12}(2150)$ $N\Delta$ dibaryon signal in the 
Dalitz plot of $M^{~2}_{d\pi^+}$ vs. $M^{~2}_{d\pi^-}$ from a preliminary 
report on $\gamma d\to d\pi^+\pi^-$ measurements in the g13 experiment 
(CLAS Collaboration) at JLab \cite{CLAS15}. Right: The $pn\to d\pi^0\pi^0$ 
WASA-at-COSY $M_{d\pi}$ distribution \cite{COSY11} and, in solid lines, 
as calculated \cite{PK16} for two input parametrizations of ${\cal D}_{12}
(2150)$. The dot-dashed line gives the $\pi{\cal D}_{12}(2150)$ contribution 
to the two-body decay of ${\cal D}_{03}(2380)$, and the dashed line gives 
a scalar-isoscalar $\sigma$-meson emission contribution.} 
\label{fig:CLAS15} 
\end{figure} 

The relevance of the ${\cal D}_{12}(2150)$ $N\Delta$ dibaryon to the physics 
of the ${\cal D}_{03}(2380)$ $\Delta\Delta$ dibaryon is demonstrated 
in Fig.~\ref{fig:CLAS15} by showing, on the left panel, a $d\pi^{\pm}$ 
invariant-mass correlation near the $N\Delta$ threshold as deduced from 
preliminary CLAS data on the $\gamma d\to d\pi^+\pi^-$ reaction \cite{CLAS15} 
and, on the right panel, a $d\pi$ invariant-mass distribution peaking near 
the $N\Delta$ threshold as deduced from the WASA-at-COSY $pn\to d\pi^0\pi^0$ 
reaction by which the ${\cal D}_{03}(2380)$ was discovered \cite{COSY11}. 
These preliminary CLAS data for $\gamma d\to d\pi^+\pi^-$ suggest 
a subthreshold ${\cal D}_{12}(2150)$ dibaryon with mass 2115$\pm$10~MeV 
and width 125$\pm$25~MeV \cite{CLAS15}, consistently with past deductions. 
The peaking of the $d\pi$ invariant-mass distribution in the $pn\to d\pi^0
\pi^0$ reaction essentially at this ${\cal D}_{12}(2150)$ mass value suggests 
that the two-body decay modes of ${\cal D}_{03}(2380)$ are almost saturated 
by the $\pi{\cal D}_{12}(2150)$ decay mode, as reflected in the calculation 
\cite{PK16} depicted in the right panel. 

The success of {\it hadronic} model calculations \cite{gg13,gg14} to reproduce 
such ${\cal D}_{IS}$ dibaryon signals is consistent with the failure of recent 
quark-based calculations \cite{Lee15} to find tightly bound {\it hexaquarks} 
by using realistic color-spin hyperfine and color confinement quark-quark 
interactions. An hexaquark with quantum numbers identical to those of 
${\cal D}_{03}$ lies at least 150~MeV above the $\Delta\Delta$ threshold, 
and this gap gets larger for other hexaquark candidates; a similar conclusion 
also holds for Jaffe's ${\cal S}=-2$ $I(J^P)=0(0^+)$ ${\cal H}$ hexaquark 
\cite{Lee16}. This means that the proper degrees of freedom in the case 
of nonstrange dibaryons are nucleons, pions and $\Delta$ baryons, and that 
physical thresholds and $p$-wave pions must be realistically incorporated 
in future considerations of such dibaryons. 

\subsection{Strange dibaryons} 
\label{sec:S=-1}  

\begin{figure}[htb] 
\centering 
\includegraphics[width=0.44\textwidth]{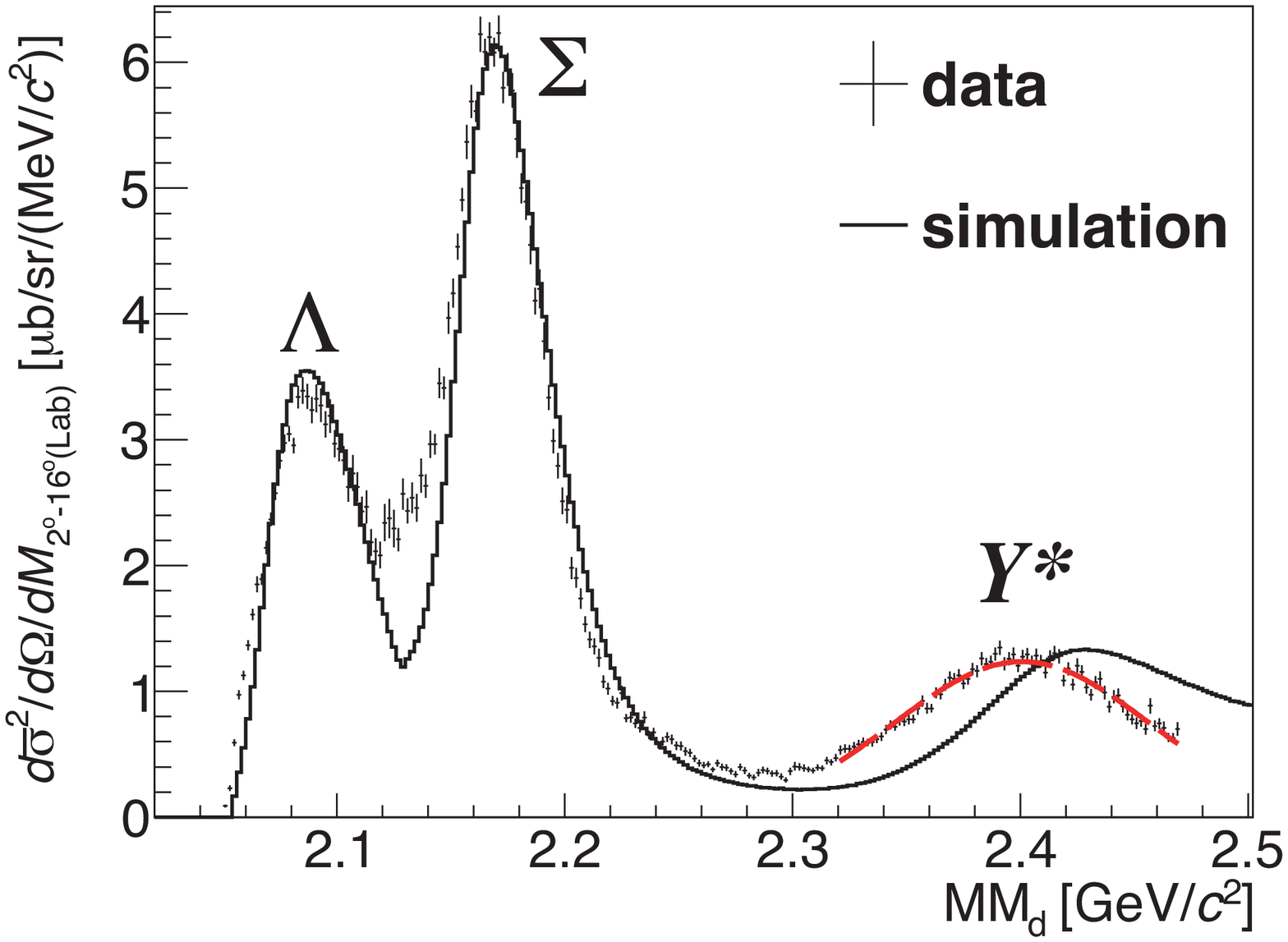} 
\includegraphics[width=0.53\textwidth]{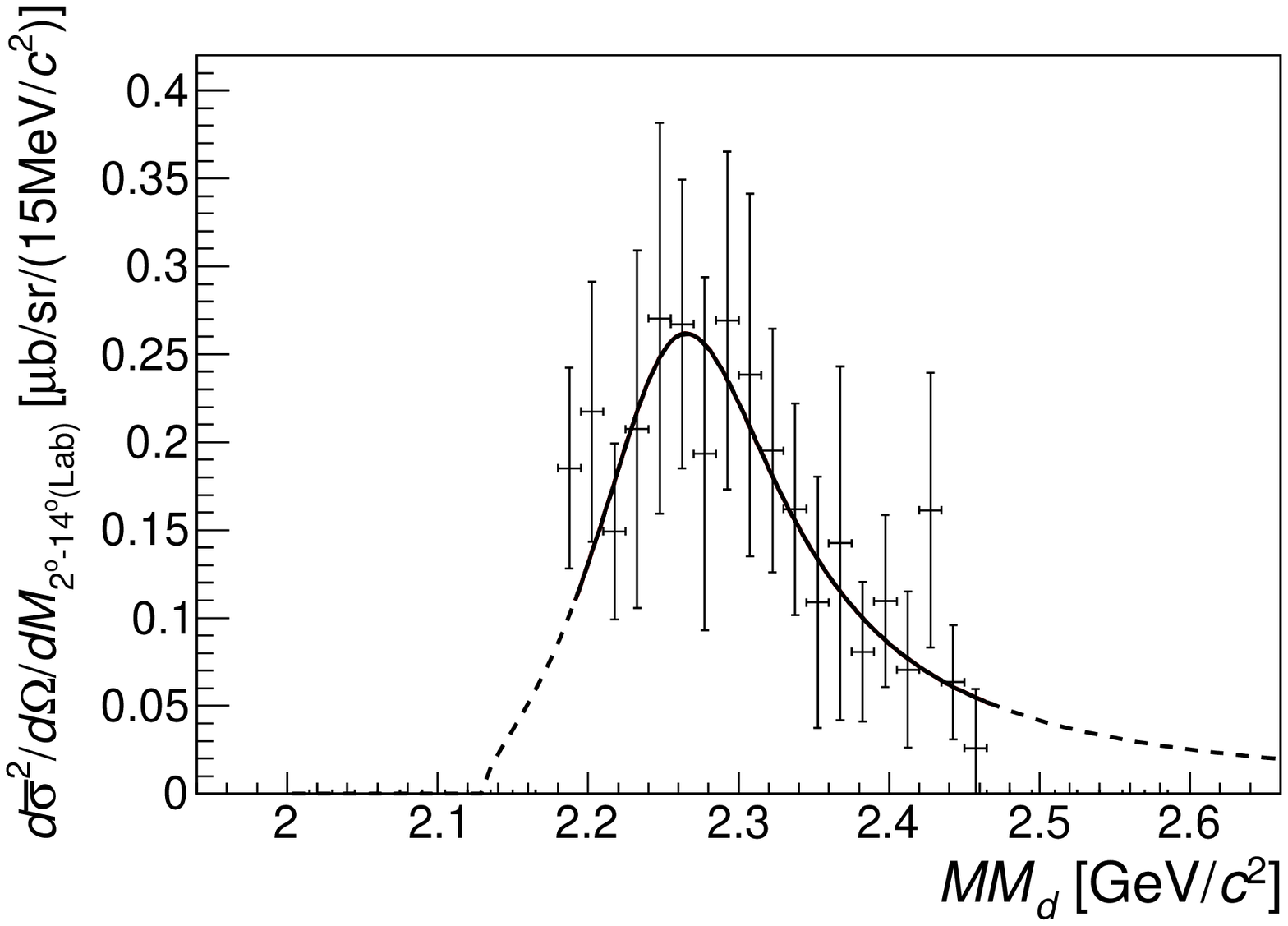} 
\caption{J-PARC E27 $d(\pi^+,K^+)X$ missing-mass spectra at 
$p_{\pi^+}=1.69$~GeV/c. Left: small-angle $K^+$ inclusive quasi-free 
spectrum \cite{E27a}. Right: $\Sigma^0 p$ decay branch of a $pp$ 
coincidence spectrum \cite{E27b}.} 
\label{fig:Y*N} 
\end{figure}

Following recent searches for a ${\bar K}NN$ $I(J^P)$=$\frac{1}{2}(0^-)$ 
quasibound state (loosely termed $K^-pp$) in Frascati \cite{finuda13,kloe16}, 
SPring-8 \cite{leps14} and GSI \cite{gsi14a,gsi15}, Iwasaki reported in 
MESON2016 on dibaryon candidates from J-PARC Experiments E27 \cite{E27a,E27b} 
and E15 \cite{E15a,E15b}, with binding energies given by 
\begin{equation} 
{\rm deep:}~~B_{K^-pp}({\rm E27})\approx 95~{\rm MeV}, \,\,\,\,\,\,\,\,\,
{\rm shallow:}~~B_{K^-pp}({\rm E15})\approx 15~{\rm MeV}, 
\label{eq:JPARC} 
\end{equation} 
relative to the $K^-pp$ threshold at 2370~MeV. To understand the possible 
origin of such radically different ${\cal S}=-1$ dibaryon candidates, it 
is instructive to look at the E27 $d(\pi^+,K^+)X$ small-angle missing-mass 
spectrum, Fig.~\ref{fig:Y*N}(left), which indicates $\approx$22~MeV 
attractive shift of the $Y^{\ast}(1385+1405)$ unresolved quasi-free peak, 
consistently with the attraction calculated in the $I(J^P)$=$\frac{1}{2}(0^-)$ 
$\Lambda(1405)N$ $s$-wave channel that overlaps substantially with $K^-pp$ 
\cite{Oka11}. Chirally motivated $K^-pp$ calculations also suggest binding 
of order 20 MeV, as reviewed in Ref.~\cite{Gal13}, in rough agreement with 
the E15 $^3{\rm He}(K^-,n)X$ near-threshold signal but not with the E27 
deeply-bound signal shown in Fig.~\ref{fig:Y*N}(right). The relatively 
shallow $K^-pp$ binding persists in three-body calculations upon including 
the $\pi\Lambda N$ and $\pi\Sigma N$ lower-mass channels \cite{RS14} which 
play only a secondary role in binding $\bar K$ mesons. 

The $\pi\Lambda N$--$\pi\Sigma N$ system, however, may benefit from sizable 
meson-baryon $p$-wave interactions, in terms of $\Delta(1232)\to\pi N$ and 
$\Sigma(1385)\to\pi\Lambda$--$\pi\Sigma$ strong-decay form factors, by 
aligning isospin and angular momentum to total $I(J^P)$=$\frac{3}{2}(2^+)$. 
Such a pion assisted dibaryon was studied in Ref.~\cite{gg13a} by solving $\pi 
YN$ coupled-channel Faddeev equations, thereby predicting a new ${\cal S}=-1$ 
dibaryon resonance ${\cal Y}_{\frac{3}{2}2^+}$ slightly below the $\pi\Sigma 
N$ threshold ($\sqrt{s_{\rm th}}\approx 2270$~MeV). Adding a $\bar KNN$ 
channel hardly matters, since its leading $^3S_1$ $NN$ component is Pauli 
forbidden. The E27 deeply bound broad signal at $\sqrt{s}\sim 2275$~MeV 
shown in Fig.~\ref{fig:Y*N}(right) may then correspond to the production of 
such ${\cal Y}^{+}_{\frac{3}{2} 2^+}$ in $\pi^{+}+d\to{\cal Y}^{+}+K^{+}$, 
followed by its decay to $\Sigma^{0}+p$ \cite{Nagae16}. We note that the 
${\cal S}=-1$ ${\cal Y}_{\frac{3}{2}2^+}(2275)$ dibaryon may have good overlap 
with $^5S_2$, $I=\frac{3}{2}$ $\Sigma(1385)N$ and $\Delta(1232)Y$ dibaryon 
configurations, the lowest threshold of which, that of $\Sigma(1385)N$, 
is only $\sim$50~MeV above the $\pi\Sigma N$ threshold. 

Other possible search reactions that are isospin selective as far as the final 
${\cal Y}\to\Sigma N$ decay is concerned are
\begin{equation} 
\pi^{\pm} + d \rightarrow {\cal Y}^{++/-} + K^{0/+}, \,\,\,\,\,\,\,\,\,\,\,\,
p + p \rightarrow {\cal Y}^{++} + K^0,  
\label{eq:Y++} 
\end{equation}  
in which the produced dibaryon ${\cal Y}$ decays to a $\Sigma N$ final charge 
state which is uniquely $I=\frac{3}{2}$, viz. ${\cal Y}^{++/-}\to\Sigma^{\pm} 
+ p(n)$. The $pp$ reaction has been reported by the HADES Collaboration 
at GSI \cite{gsi14b}, finding no $\cal Y$ dibaryon signal. It is not clear 
whether the $pp$ experiments were able to resolve as small cross sections 
as 0.1~$\mu$b or less that are expected in production of $\cal Y$ dibaryon 
candidates \cite{gsi15}.

\subsection{Charmed dibaryons}
\label{sec:C=+1}

Charmed, ${\cal C}=+1$ dibaryons have also been predicted: 
(i) a $I(J^P)$=$\frac{1}{2}(0^-)$ dynamically generated $DNN$ quasibound 
state at 3.5~GeV \cite{Oset12} reminiscent of the ${\cal S}=-1$ $K^-pp$; 
and (ii) a $I(J^P)$=$\frac{3}{2}(2^+)$ $\pi\Lambda_c N$ quasibound state below 
3.4~GeV \cite{GGVFC14}, analogous to the ${\cal S}=-1$ pion assisted dibaryon 
${\cal Y}_{\frac{3}{2}2^+}$. The prediction of this charmed pion assisted 
dibaryon ${\cal C}_{\frac{3}{2}2^+}(3370)$ is robust, since it depends little 
on the unknown $^3S_1$ $\Lambda_c N$ interaction. The ${\cal C}_{\frac{3}{2}
2^+}(3370)$ is likely to be the {\it lowest lying} charmed dibaryon. It could 
be searched with proton beams at GSI, and with pion beams in the high-momentum 
hadron beam line extension approved at J-PARC, viz. 
\begin{equation} 
(p + p)~~{\rm or}~~(\pi^+ + d)~\rightarrow~{\cal C}^{+++} + D^-,
\,\,\,\,\,\,\,\,\,\, 
{\cal C}^{+++}\rightarrow \Sigma_c^{++}(2455) + p. 
\label{eq:C+++} 
\end{equation}

\section{Meson-nuclear bound states} 
\label{sec:mesonBS} 

\begin{figure}[htb] 
\centering 
\includegraphics[width=0.48\textwidth]{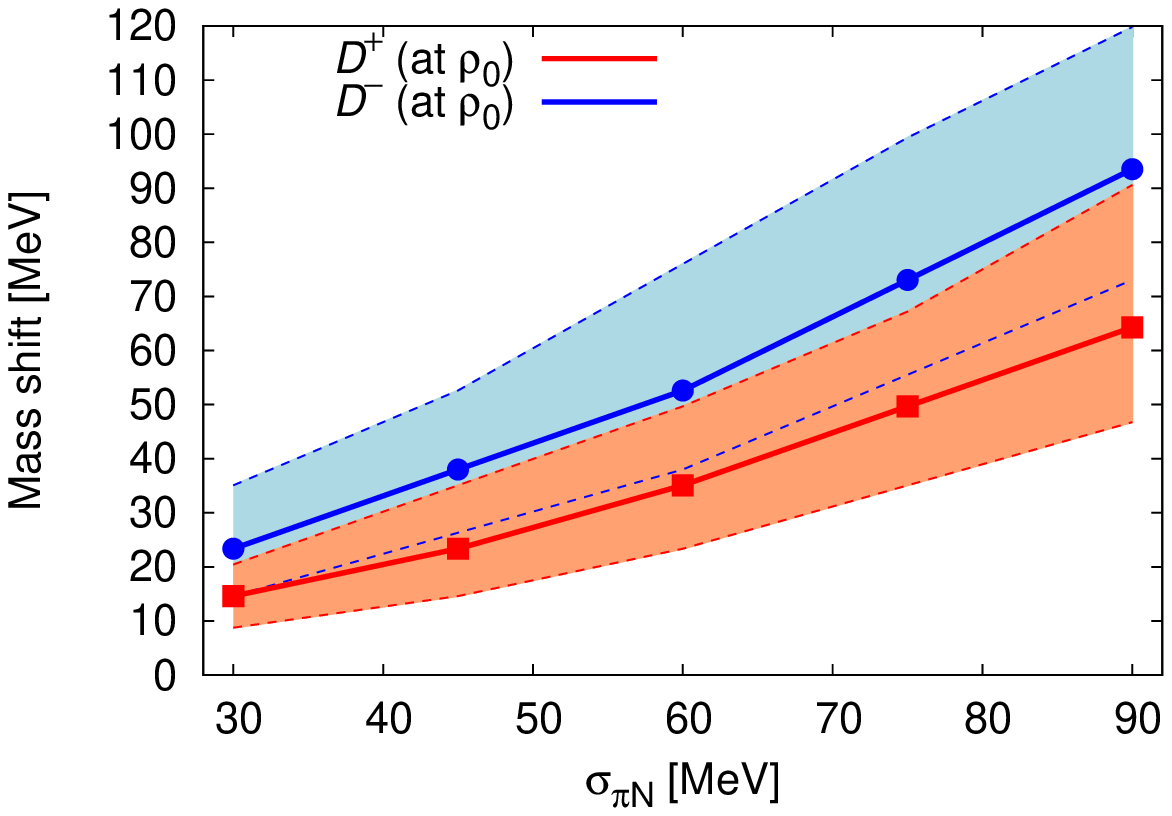} 
\includegraphics[width=0.48\textwidth]{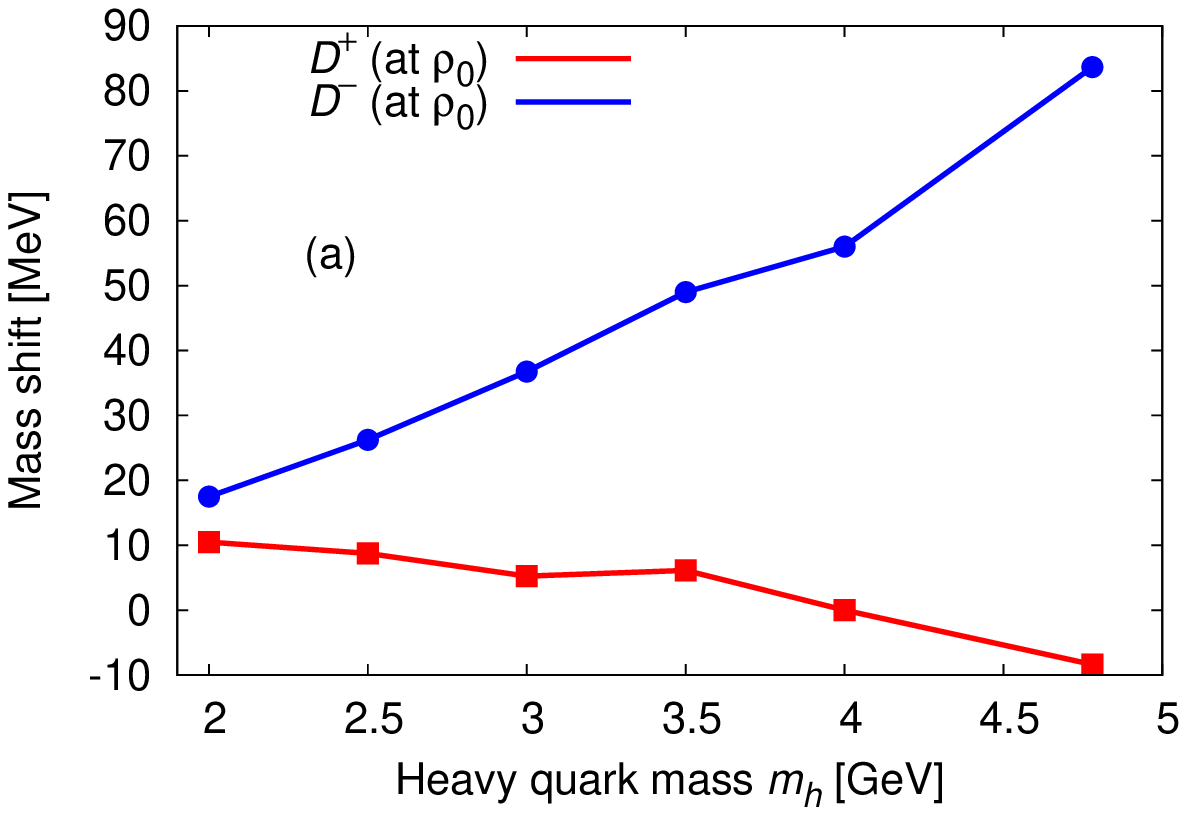} 
\caption{$D$ meson mass shift in nuclear matter \cite{Oka16} vs. 
$\sigma_{\pi N}$ (left) and vs. the heavy quark mass m$_h$ (right).} 
\label{fig:Dmass} 
\end{figure}

No meson-nuclear bound states have been firmly established so far. For $K^-$ 
mesons, extrapolating from kaonic atoms \cite{FG12} it is widely accepted 
that broad quasibound states exist \cite{GM12}. $K^+$ mesons, in contrast, 
experience repulson in dense (nuclear) matter. This is naively explained by 
a mean-field treatment of $K^-\equiv s{\bar u}$ and $K^+\equiv {\bar s}u$ 
mesons, arguing that a nonstrange antiquark/quark induces attraction/repulsion 
in dense matter. In the charmed sector, one would then expect {\it attraction} 
for $D^+\equiv c{\bar d}$ and repulsion for $D^-\equiv{\bar c}d$ mesons. 
This is not borne out in a recent QCD sum-rule calculation, showing in 
Fig.~\ref{fig:Dmass}(left) a {\it repulsive} mass shift in dense matter, 
as a function of the assumed value of the $\pi N$ $\sigma$ term, for 
{\it both} $D^{\pm}$ mesons \cite{Oka16}. Fig.~\ref{fig:Dmass}(right) shows 
that an attractive $D^+$ mass shift is possible in principle, but only for 
unrealistically high values of the heavy-quark mass $m_h$. This result has 
also been explained in Ref.~\cite{Hrada16} using a constituent quark picture. 

Turning to nonstrange and noncharmed meson-nuclear interactions discussed in 
MESON2016, Nanova reviewed recent $\omega$ and $\eta'$ nuclear photoproduction 
experiments by the CBELSA/TAPS Collaboration which study the meson momentum 
dependence of the extracted meson-nucleus optical potential, suggesting that 
while the $\omega$-nucleus potential is too absorptive to observe distinct 
quasibound states, the $\eta'$-nucleus potential is weakly absorptive 
\cite{Friedrich16} and sufficiently attractive \cite{Nanova16} to motivate 
searches for $\eta'$-mesic nuclear states. Ongoing searches in $^{12}$C$(p,d)$ 
at GSI were discussed in Tanaka's talk. However, with $p_{\eta'}$ centered 
about $\sim$1~GeV in the ELSA experiments, the optical potential derived for 
$\eta'$ at rest depends on extrapolation from the lowest available value 
$p_{\eta'}\approx 275$~MeV/c down to threshold, where COSY-11 data on 
near-threshold meson production in $pp$ collisions indicate a rather strong 
$\eta p$ attraction that is likely to support $\eta$-mesic nuclear states 
and a much weaker $\eta' p$ interaction \cite{Moskal00}, for the real part 
of which only a limit consistent with zero can be placed \cite{Moskal14}. 
Citing from Wilkin's talk: ``I would not put any money on bound $\eta'$ in 
nuclei!". Within a QCD-inspired $\eta -\eta'$ mixing model \cite{Bass14}, 
$\eta'$-nuclear attraction of roughly $-$40~MeV at saturation density as 
derived in the ELSA experiment \cite{Nanova16} implies about $-$90~MeV for 
$\eta$ attraction in nuclear matter, commensurate with the attraction expected 
in the $\eta N$ interaction model GW considered below. For these and for other 
reasons specified below, the following discussion is limited to $\eta$-nuclear 
quasibound states. 

\begin{figure}[htb] 
\centering 
\includegraphics[width=0.48\textwidth,height=4.5cm]{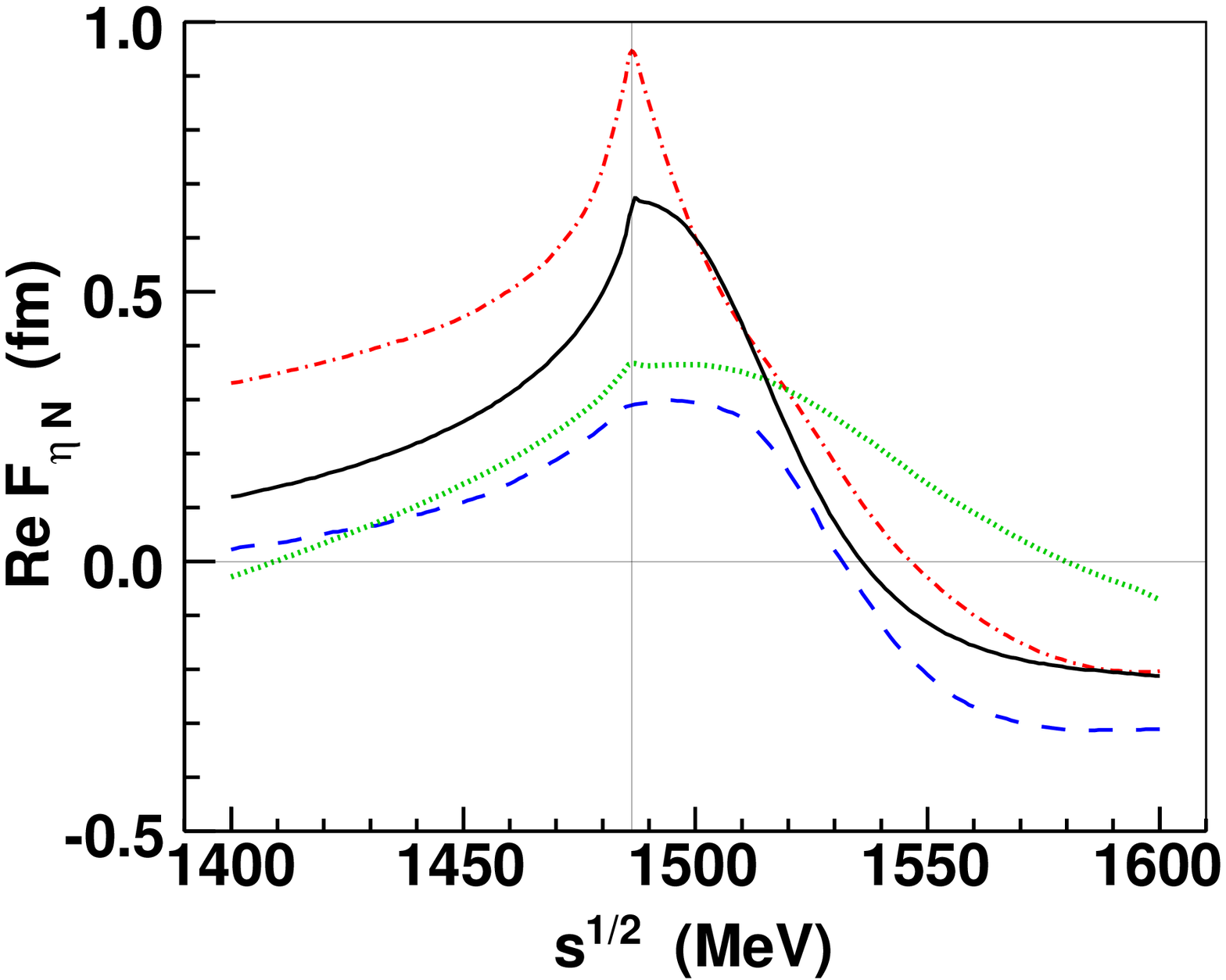} 
\includegraphics[width=0.48\textwidth,height=4.5cm]{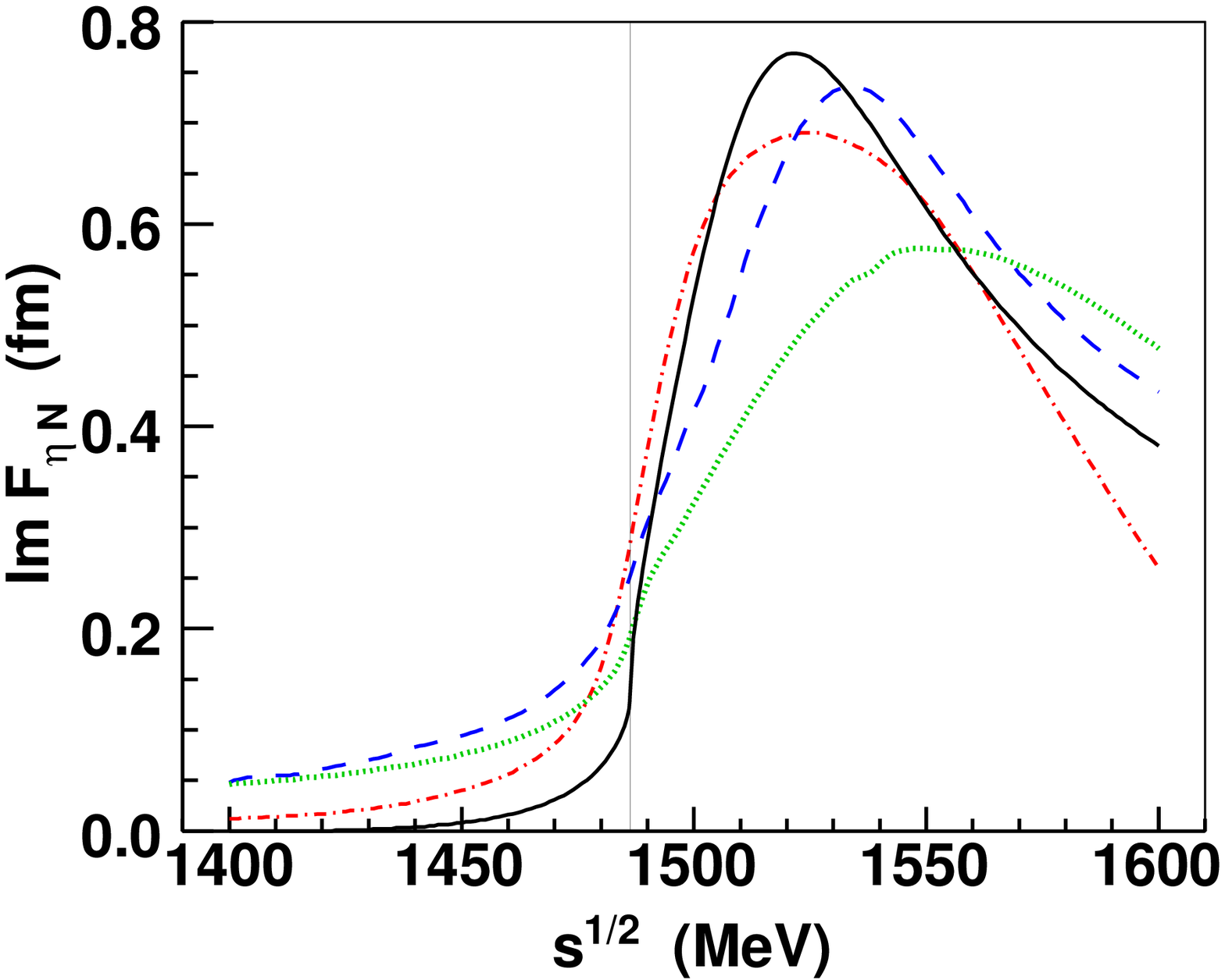} 
\caption{Real (left) and imaginary (right) parts of the $\eta N$ $s$-wave cm 
scattering amplitude $F_{\eta N}(\sqrt s)$, compiled in Ref.~\cite{Gal14} from 
several $N^{\ast}$(1535) resonance coupled-channel models, in decreasing order 
of Re$\;a_{\eta N}$: GW \cite{GW05}, CS \cite{CS13}, MBM \cite{MBM12} and 
IOV \cite{IOV02}. The $\eta N$ threshold is marked by a thin vertical line.}  
\label{fig:etaN} 
\end{figure} 

The $\eta N$ near-threshold dynamics is governed by the $N^{\ast}$(1535) 
resonance, introducing thereby appreciable model dependence in coupled-channel 
calculations of the $s$-wave scattering amplitude $F_{\eta N}(\sqrt{s})$, 
as seen in Fig.~\ref{fig:etaN}. Owing to the nearby $N^{\ast}$(1535), both 
Re$\,F_{\eta N}$ and Im$\,F_{\eta N}$ decrease in all models steadily below 
threshold, which is where bound states are calculated. This decrease persists 
also in in-medium extensions $F_{\eta N}(\sqrt{s},\rho)$ of the $\eta N$ 
scattering amplitude, suggesting that $\eta$-nuclear states are narrow. I know 
of no similar mechanism that would suggest as narrow $\eta'$-nuclear states. 

\begin{figure}[htb]
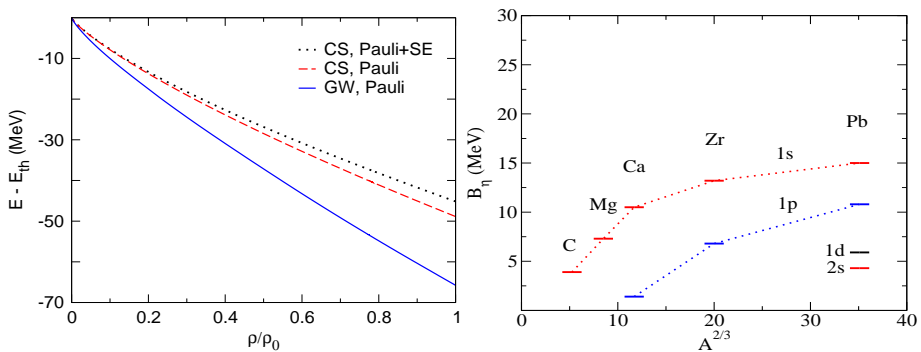
 
\centering 
\includegraphics[width=0.42\textwidth,height=4.4cm]{srhocs3.eps} 
\includegraphics[width=0.42\textwidth,height=4.5cm]{beta-CS.eps} 
\caption{Left: Subthreshold energies probed in the $1s_{\eta}$-$^{40}$Ca 
bound state as a function of nuclear density, calculated self-consistently 
for in-medium $\eta N$ scattering amplitudes in models GW and CS used in 
Ref.~\cite{FGM13}. Right: $\eta$-nuclear spectra \cite{CFGM14} calculated 
self-consistently using the in-medium CS model NLO30$_{\eta}$ \cite{CS13}.} 
\label{fig:etaNucleus} 
\end{figure} 

The subthreshold energy $\sqrt{s}$ and the nuclear density $\rho$, 
both serving in bound state calculations as arguments of the in-medium 
meson-nucleon scattering amplitude $F_{mN}(\sqrt{s},\rho)$, are tightly 
correlated as demonstrated in Fig.~\ref{fig:etaNucleus}(left) within 
a particular $\eta$-nucleus calculation. This correlation imposes 
a self-consistent procedure in bound state calulations \cite{Gal14}, 
as discussed here by Mare\v{s}. 

A chart of $\eta$-nuclear bound states calculated self-consistently in the 
CS model is shown in Fig.~\ref{fig:etaNucleus}(right). Since Im$\,F_{\eta N}
(\sqrt{s})$ is particularly small in model CS below threshold, see 
Fig.~\ref{fig:etaN}(right), the resulting $\eta$-nuclear widths are just 
a few MeV, and only somewhat larger in model GW. Bound states should 
definitely exist in $^{12}$C and beyond, and beginning in $^6$Li in model 
GW which according to Fig.~\ref{fig:etaN}(left) provides the strongest 
$\eta N$ attraction among the four models exhibited, Few-body calculations 
have also been reported recently using the $\eta N$ interaction models GW and 
CS. No bound state was found for $\eta d$ and for $\eta \, ^3$He \cite{BFG15}. 
Calculations are underway for $\eta \, ^4$He.

\section{Summary and outlook} 
\label{sec6} 

Several topics discussed in MESON2016 were picked up selectively in these 
Concluding Remarks, the common grounds of which is the rich spectroscopic 
variety that remains largely to be uncovered in hadronic systems. The 
impression gained at this Conference is that no consensus has been reached 
on the hidden-charm structures observed recently for mesons and for baryons 
in the energy range 2--5 GeV. In the absence of compelling arguments, 
or calculations classifying these in terms of genuine tetraquarks and 
pentaquarks, the only logical conclusion is that of hadronic-molecule 
underlying structure. For dibaryons too, highlighting recent experimental 
results from COSY (nonstrange) and J-PARC (strange), the evidence points to 
hadronic structure. Finally, we focused attention to the possible existence 
of observable $\eta$-nuclear quasibound states, which could be explored at GSI 
using $(p,d)$ and in J-PARC using the $(\pi^+,p)$ reaction on nuclear targets.

\begin{acknowledgement}

I would like to thank the Organizers of MESON2016 for trusting me in this 
unthankful job. 

\end{acknowledgement}

\end{document}